\documentclass[prl,twocolumn,amsmath,amssymb,aps,floatfix]{revtex4-1}

\usepackage{dcolumn}
\usepackage{bm}
\usepackage{hyperref}
\usepackage[usenames,dvipsnames]{xcolor}
\usepackage[normalem]{ulem}
\usepackage[utf8]{inputenc}
\usepackage{amsmath}
\usepackage{amsthm}
\usepackage{amsmath}
\usepackage{amsfonts}
\usepackage{blindtext}
\usepackage[utf8]{inputenc}
\usepackage[T1]{fontenc}
\usepackage{graphicx}
\usepackage{float}
\usepackage{wrapfig}
\usepackage{bm}
\usepackage{braket}
\usepackage{grffile}
\usepackage[usenames,dvipsnames]{xcolor}
\newcommand{\beq}{\begin{equation}}
\newcommand{\eeq}{\end{equation}}

\newcommand{\bk}{\boldsymbol{k}}
\newcommand{\bq}{\boldsymbol{q}}
\newcommand{\bp}{\boldsymbol{p}}
\newcommand{\bx}{\boldsymbol{x}}

\newcommand{\bu}{\boldsymbol{u}}

\newcommand{\bw}{\boldsymbol{\omega}}
\newcommand{\bhu}{\boldsymbol{\hat{u}}}

\begin{document}

\title{Phase transitions and flux-loop metastable states in rotating turbulence} 

\author{P. Clark Di Leoni$^1$, A. Alexakis$^2$, L. Biferale$^3$ and M. Buzzicotti$^3$}
\affiliation{$^1$Department of Mechanical Engineering, Johns Hopkins
University, Baltimore, Maryland 21218, USA.}
\affiliation{$^2$Laboratoire de Physique de l'\'Ecole Normale Sup\'erieure, CNRS,
    PSL Research University, Sorbonne Universit\'e, Universit\'e de Paris,
    F-75005 Paris, France.}
\affiliation{$^3$Dept. Physics and INFN, University of Rome ``Tor Vergata'', Italy.}

\date{\today}

\begin{abstract}

By using direct numerical simulations of up to a record resolution of
512x512x32768 grid points we discover the existence of a new metastable
out-of-equilibrium state in rotating turbulence.  We scan the phase
space by varying both the rotation rate (proportional to the inverse of
the Rossby number, $Ro$) and the dimensionless aspect ratio,
$\lambda=H/L$, where $L$ and $H$ are the sizes of the domain
perpendicular and parallel to the direction of rotation, respectively.  We show the
existence of three turbulent phases. For small $Ro$ but finite
$\lambda$, we have a split cascade where the injected  energy is
transferred to both large and small scales.  For large $\lambda$ and
finite $Ro$ there is no inverse cascade and the energy is
transferred forward in Fourier space only. Surprisingly, between these
two regimes, a third phase is observed as reported here for the first
time.  Consequently, for certain intervals of $Ro$ and $\lambda$,  energy is no
longer accumulated at arbitrarily large scales, rather it stops at some
characteristic intermediate length-scales from where it is then redistributed
forward in Fourier space, leading to a flux-loop mechanism where the
flow is out of equilibrium with vanishing net flux, and non-vanishing
heterochiral and homochiral sub-fluxes.  The system is further
characterized by the presence of metastability and critical slowing
down, explaining why previous experiments and numerical simulations were
not able to detect this phenomenon, requiring extremely long observation
time and huge computational resources. 

\end {abstract}
\maketitle

\begin{figure*}
    \centering
    \includegraphics[width=0.32\textwidth]{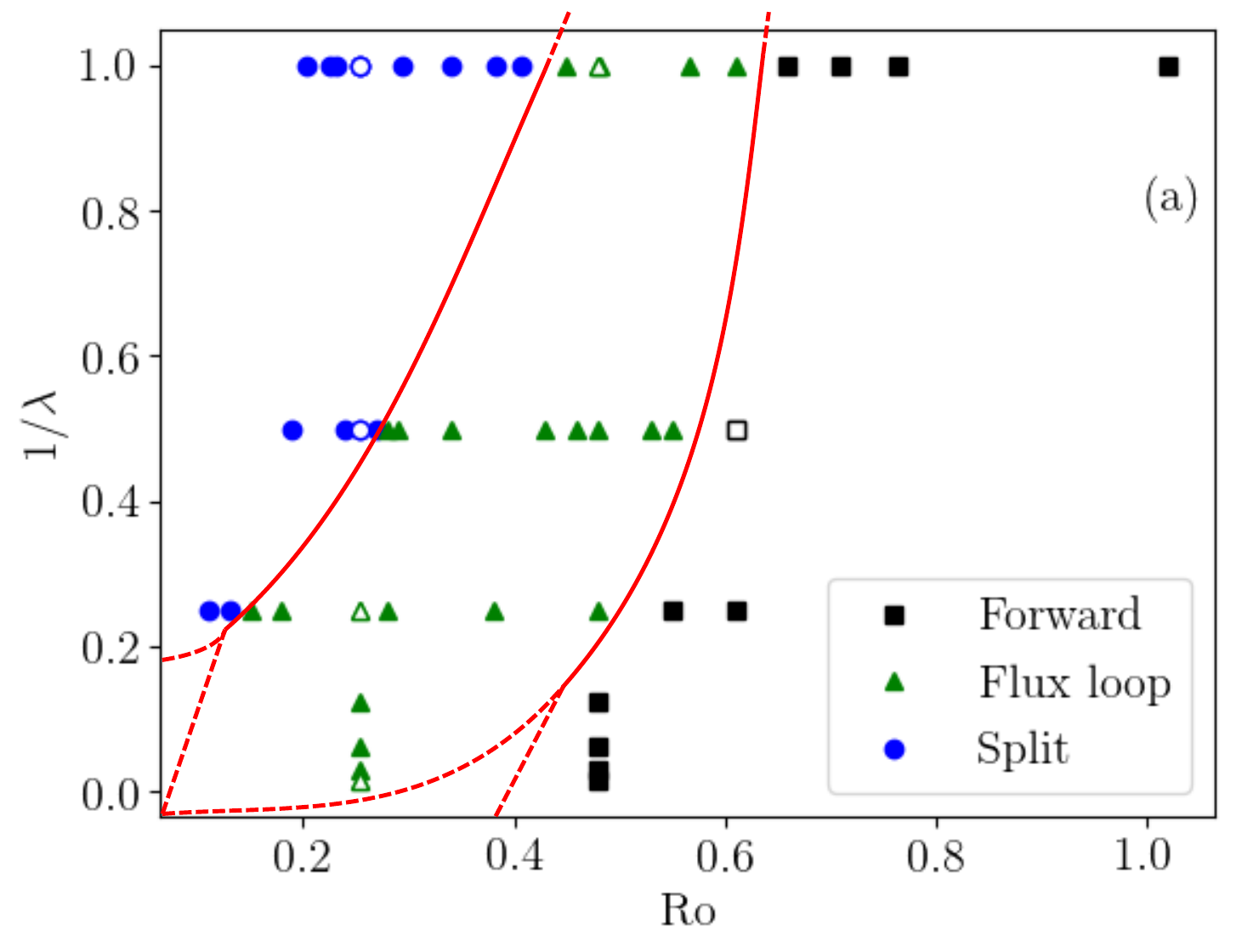} 
    \includegraphics[width=0.32\textwidth]{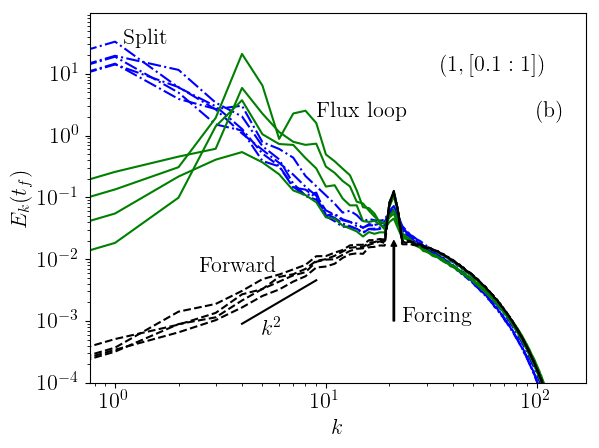}   
    \includegraphics[width=0.32\textwidth]{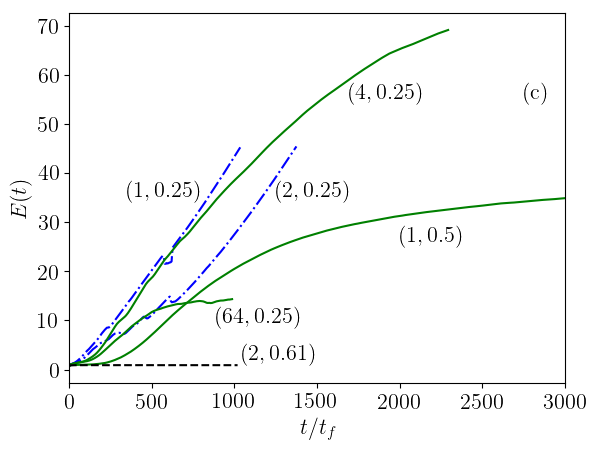}   
    \caption{(a): $(\lambda,Ro)$ Phase space. Different symbols represent
    the three macroscopic turbulent cascade phases: forward, flux-loop
    and split. Red continuos line are a guide to the eye to distinguish
    the three phases. Red dotted-lines represents possible asymptotics
    behaviour in the limit $Ro \to 0$ and $\lambda \to \infty$ (see
    discussion in the text). (b): Instantaneous energy spectra
    for fixed aspect ratio, $\lambda=1$ at different Rossby numbers $
    0.1 \le  Ro \le 1$. For flux-loop and direct cascade cases the spectra are plotted in the 
    stationary regime, for the split-cascade regime we used the final time when we stopped the 
    simulation. (c): time evolution of the total energy, $E(t)$,
    for some characteristic $(\lambda,Ro)$  values (represented with
    empty symbols in panel (a). Line colours distinguish the
    three phases following the same color code of symbols in panel (a).}
    \label{spectra}
\end{figure*}

{\sc Introduction.} Statistical
systems can develop critical behaviour, where abrupt macroscopic
changes happen when varying some  control parameter, like temperature or 
magnetic field \cite{goldenfeld2018lectures}.  Averaged quantities
can show discontinuous or continuous variations across the critical
lines/points in the phase space where the transition occurs.  In many
cases, experimental and numerical realizations are affected by long
transients that are generated due to the presence of metastable states corresponding to
local minima of the (free) energy.  There have been many attempts to
transfer such descriptions to  out-of-equilibrium systems quantitatively
\cite{zwanzig2001nonequilibrium,berthier2011theoretical}. The critical behaviour of stationary systems in the presence of energy injection mechanisms, dissipation and
non-vanishing fluxes remains a major topic of current research in fluid-dynamics, granular and active matter and lacks
 systematic theoretical understanding \cite{marchetti2013hydrodynamics,alexakis2018cascades,aranson2006patterns}.  A
paradigmatic example of (phase) transition
 is the sudden jump from laminar to turbulent dynamics that is observed in  Poiseuille and Couette flows when changing the forcing intensity \citep{lemoult2016directed,
chantry2017universal, moxey2010distinct}. In this paper, we investigate the important case of rotating turbulence where the control parameter, given by the intensity of the Coriolis force,  affects the symmetries of the macroscopic flow but not the energy injection rate \cite{davidson2013turbulence,campagne2014direct,pouquet2013geophysical,yarom2014experimental,sagaut2008homogeneous}. 
In this set-up, it is known that for
sufficiently weak rotation rate, $\Omega$,  the system behaves as 3D
homogeneous and isotropic turbulence transferring energy to small scales
only ({\it forward energy cascade}), while for  $\Omega$ above a critical
value, $\Omega_c$, 3D fluctuations are sufficiently suppressed, the flow
becomes quasi-two dimensional and energy is transferred with a {\it
split cascade} to large scales also
\cite{smith1999transfer,alexakis2018cascades}.  As  as the domain size
$H$ in the direction of rotation becomes larger, $\Omega_c$ increases
\cite{deusebio2014dimensional,pestana2019regime,pestana_rossby-number_2020}.
{Arguments based on wave-turbulence theory suggest that the inverse flux
vanishes in the infinite volume limit, predicting that $\lim_{H \to \infty} \Omega_c \to
\infty$ \cite{galtier2003weak}.} {More recently, considering an
asymptotic form of the governing equations for large $\Omega$, it was
shown that $\Omega_c\propto H$ \cite{van2019critical}.} 
However, for finite $\Omega$ we do not know the precise functional dependence of the boundary $\Omega_c(H)$ nor the nature of the transition. 
Despite of the importance of such
questions for geophysical and engineering applications we still do not have a satisfying understanding of any of them.   While trying to
address these problems, we discovered that the physics is even richer due to  the existence of a new region in the
$(\lambda,Ro)$ phase-space, where turbulence develops a third,
non-trivial, macroscopic out-of-equilibrium state, characterized by a
{\it flux-loop cascade}, where the flow organises to spontaneously contain the
tendency to condensate energy in larger and larger scales, entering a
stationary regime and  producing a quasi-ordered array of turbulent
columnar vortices, akin to a vortex crystal \cite{aref_vortex_2002}.  The
aim of this letter is to report about this new state of turbulent rotating flows, characterizing
its peculiar statistical and dynamical properties, including
metastability and  critical slowing down  {as well as commenting about
its potential importance for other turbulent realizations. }

{\sc Setup.} We begin by considering the flow in a rectangular periodic
domain with aspect ratio $\lambda\equiv H/L $  and dimensions {$2\pi L \times 2\pi L \times  2 \pi H$} in a
rotating reference frame where the rotation is along the direction with
dimension $2 \pi H$.  
The governing equations for the
incompressible velocity field, ${\bf u}$, can be written as

\begin{equation}
    \partial_t {\bf u} + {\bf u \cdot \nabla u} + 2{\bf \Omega  \times u
    } =-\nabla P + \nu \Delta {\bf u} + {\bf f} \label{NSR}
\end{equation}
where $\nu$ is the kinematic viscosity, ${\bf f}$ is an external forcing
and  $2 {\bf \Omega \times u}$ is the Coriolis force produced by the frame rotating with intensity $\Omega$.   We choose the forcing ${\bf f}$ to be
Gaussian and delta-correlated in time, acting on a spherical Fourier
shell with $|{\bf k }|\in [k_f,k_f + 2]$ and with amplitude $f_0 =
f_1/\lambda$,  so that the total
injection rate $\epsilon = \langle {\bf f} \cdot {\bf u} \rangle$
remains fixed, where $\langle \bullet \rangle$ means an average over the
forcing realization or on the whole fluid volume.  In order to reduce viscous effects, most of the results shown have been obtained by adopting a modified hyper-viscosity (see later and SM for details).  In the
present work, we keep $k_f L=20$ fixed, using  $\lambda$
as  control parameter.  
Besides $\lambda$ the other two
non-dimensional quantities are given by Reynolds,
${Re}=\epsilon^{1/3} k_f^{-4/3} \nu^{-1} $ and Rossby,
${Ro}={\epsilon^{1/3}k_f^{2/3}}\Omega^{-1}$, numbers.   The equations are solved using a parallel
pseudo spectral code (see details in \cite{biferale2016coherent}) using grids as big as $512 \times 512\times 32768$ for the largest aspect ratio $\lambda = 64$. It is important to stress that accessing high aspect ratios is key to attack the {\it infinite volume}
limit in the direction parallel to rotation and to assess potential
singular effects induced by a finite separation of the 2D plane at
$k_\parallel=0$ from the 3D modes with $k_\parallel >0$ in Fourier space
\cite{waleffe1993inertial}.
We also fixed  $L=1$ and $\epsilon=1$. 

{\sc RESULTS.} Figure~\ref{spectra}(a) summarizes the main results of our
paper,  showing the existence of three different macroscopic {\it
phases} of the rotating flow in the $(\lambda,Ro)$ space, consisting in
 (i) a pure forward cascade, (ii) a new flux-loop regime {(the choice
of name will become obvious later)} and  (iii) a split cascade regime.
In Fig.~\ref{spectra}b we show for the sub-set of simulations with
$\lambda=1$ and at various ${Ro}$ the isotropic energy spectra,
defined as $$E_k(t)=\frac12 \sum_{k\leq|\bm{k}|<k+1}|\hat{\bm
u}_{\bm{k}}(t)|^2$$ where $\hat{\bm u}_{\bm{k}}(t)$ are the Fourier
coefficients of the velocity field. 
The direct
cascade regime (black dashed lines) does not develop any large scale
fluctuations and peaks at the forcing scale. The split cascade
regime (blue dash-dotted curves) showcases both forward and inverse
cascades and the simulations are stopped when the peak at the largest horizontal scale, $k \simeq 1$, is well developed. The novelty here is given by the flux-loop
phase (solid green lines) showing an intermediate spectral behavior. In
this case,  the energy spectra are much more irregular, break
self-similarity and have their largest peak at an intermediate
wavenumber, $k\simeq 5$. Figure~\ref{spectra}(c) shows the evolution of
the total energy, $E(t) = \sum_k E_k(t)$, for some of the most
characteristic $(\lambda,Ro)$ values.  As expected, in the case with a
forward cascade we have a constant (small) total kinetic energy as all
energy input is dissipated at high wavenumbers by the viscosity. In the two
data-sets showing a split cascade regime, the energy increases constantly
as it is transferred  to  large scales without important dissipative
effects. In the three flux-loop data-sets,  the total energy saturates
for very long times, indicating that the flux to the large scales is
halted.

\begin{figure*}
    \centering
    \includegraphics[width=0.99\textwidth]{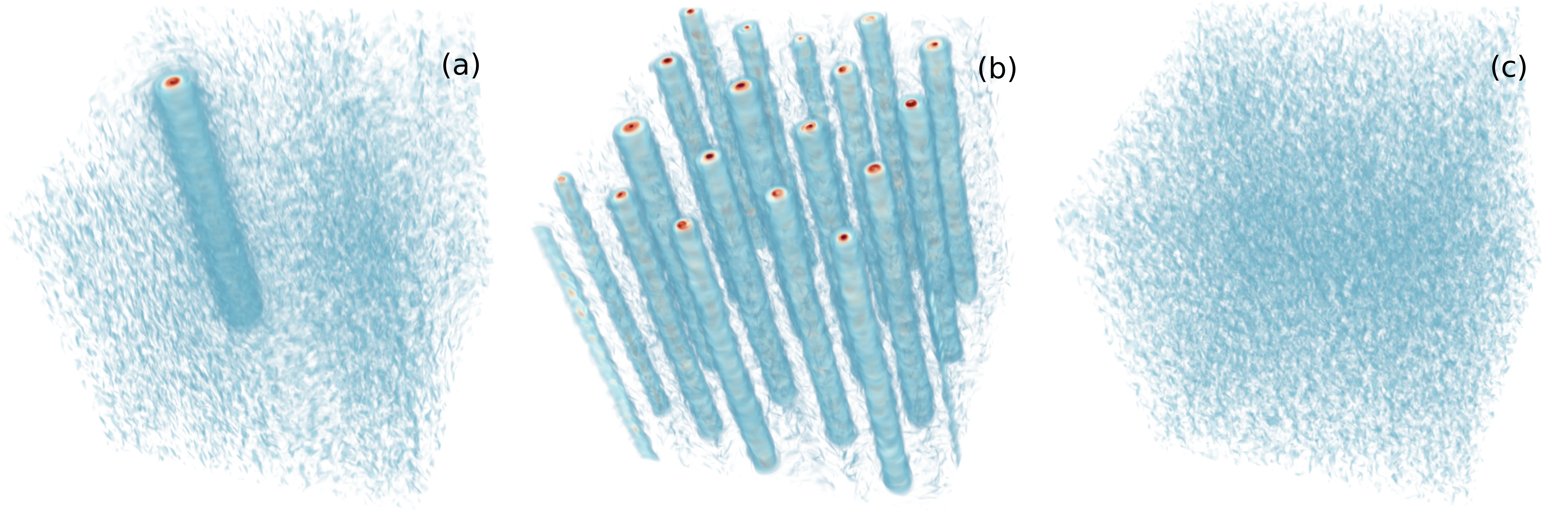}
    \includegraphics[width=0.99\textwidth]{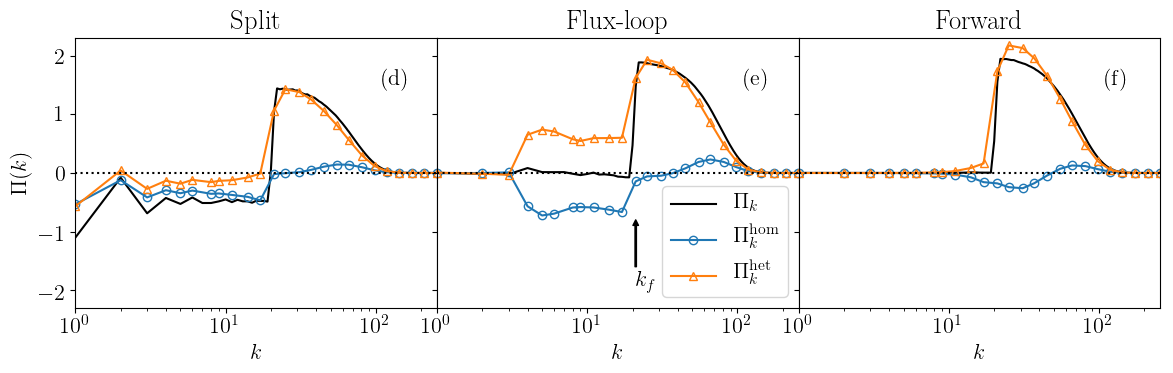} 
    \caption{(a)-(c): Visualizations of the parallel component of the
    vorticity for a split cascade case, a flux loop case, and a forward
    cascade case, respectively. (d)-(f): Total and chirally decomposed
    fluxes, for the three regimes shown in the top row.}
    \label{viz_and_flux}
\end{figure*}

In Figs.~\ref{viz_and_flux}(a)-(c) we show visualisations of the vorticity projection in the 
direction of the rotation axis 
for three characteristic data-sets representing the three different phases at late times, and in the SM we show movies comparing their time evolutions.
In the split cascade regime (a), the system  forms many co-rotating columnar
vortices which eventually merge into one.  In the forward cascade regime
(c), no large scale coherent vortical structures are formed, as
expected. In the new flux-loop regime (b), the columnar vortices form,
but do not merge and get quasi stuck in a lattice-like structure that persists
in time. Similar structures, deemed ``vortex crystals'', have been
observed in systems like 2D point vortices \cite{aref_point_1998,aref_vortex_2002b}, 2D
turbulence \cite{fine_relaxation_1995,huang1994relaxation,jimenez_spontaneous_2007},
Bose-Einstein condensates \cite{abo2001observation, newton_vortex_2009}, and even Jupiter's
atmosphere \cite{adriani_clusters_2018}. In
particular, asymmetric states where the system resembles a crystal with
defects, like what is shown in Fig.~\ref{viz_and_flux}b, have been shown to
be equilibria of co-rotating point vortex systems \cite{aref_point_1998,aref_vortex_2002b}. 
Note that in 2D randomly forced turbulence there is a symmetry between positive and negative 
vorticity. As a result, the kind of structures that we observe here are connected to
the asymmetry between co-rotating and counter rotating vortices introduced by rotation and 3D 
effects.

It is worth noting that the vortex crystal state is formed in
the absence of any large scale damping term to suppress the
inverse cascade.  The stationarity of the energy spectrum then implies
that the total inverse energy flux at $k<k_f$ has to be zero.
Nonetheless, the spectrum is far from the $E_k \propto k^2$ shape
predicted by a simple equilibrium distribution
\cite{dallas2015statistical,alexakis2019thermal}.  To resolve this
puzzle we show in Figs.~\ref{viz_and_flux}(d)-(f) the total energy flux,
$\Pi_k=- i \sum_{|\bm{k}|\le k}   \sum_{\bm{p}+\bm{q}=\bm{k}} (\hat{{\bm
u}}_{-\bm{k}} \cdot \hat{{\bm u}}_{\bm{p}}) ({\bm k} \cdot \hat{{\bm
u}}_{\bm{q}})$,  and its
exact decomposition in homochiral and heterochiral sub-components, $\Pi_k =
\Pi^{hom}_k+\Pi^{het}_k$, built in terms of Fourier triads including
modes with the same or opposite helicity signatures (see SM and
\cite{waleffe_nature_1992,waleffe1993inertial} for a discussion
about the importance of hetero- and homo-chiral properties for the energy cascade direction).
In  order to reduce fluctuations, fluxes are averaged on stationary or quasi-stationary time 
windows. As one can see, the forward cascade  (panel f) does
not show any exotic behaviour, both the total flux and it sub-components are zero in the
$k<k_f$ range
\cite{alexakis_helically_2017,sahoo_energy_2018}.  For the split cascade
phase in the range $k<k_f$ (panel d) we have the usual negative total
flux as a result of the negative contributions from the two helical sub-components (in a quasi 2D
regime helicity does not play any role and homo- and hetero-chiral
channels are expected to be identical \cite{biferale2017two,buzzicotti2018energy}).  The interesting -and non
trivial- result is shown in panel (e) where the total flux for $k<k_f$
is zero, as it must be if the statistics are stationary, but it is the result of
a balance between the forward, $\Pi_k^{het}>0$,  and the inverse, $\Pi_k^{hom}<0$ sub-fluxes contributions.
Hence the name of a flux-loop state \cite{alexakis2018cascades}.  This
highly intricate flux-loop balance is an out-of-equilibrium  effect and
has already been observed in two-dimensional but three-component flows
\cite{biferale_two-dimensional_2017} and in rotating flows with only
three-component motions \cite{buzzicotti_inverse_2018} where similar
peaked spectra were found. 

\begin{figure}
    \centering
    \includegraphics[width=0.45\textwidth]{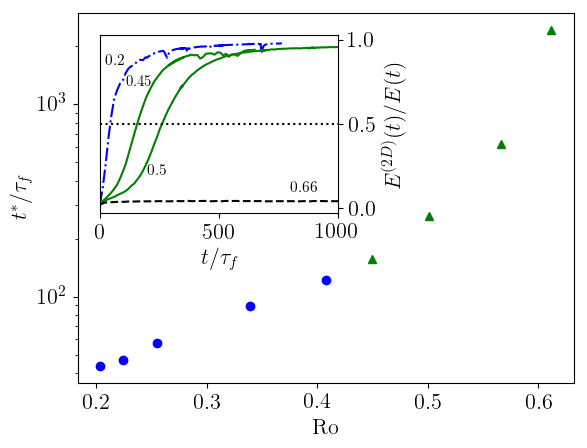}
    \caption{Time to bidimensionalization $t^*$ as a function of
    Rossby number for aspect ratio $\lambda=1$. Inset: Ratio of the
    energy in the 2D modes a function of time.}
    \label{bidim_dynamics}
\end{figure}

From  previous figures it is clear that the inverse cascade and the
flux-loop phase are the results of a competition between a tendency to
become 2D-like contrasted by some residual 3D structures that push energy
forward. It is therefore interesting to assess the dynamical effects of
these contrasting forces. To do that, we measured the typical time
it take the energy to become concentrated on the $k_\parallel
\approx 0$ 2D plane, defined as the instant of time, $t^*$, when the
ratio ${E^{(2D)}(t^*)/E(t^*)}$ hits $0.5$, where $E^{(2D)}(t)=\sum_{|{\bm
k}\cdot{\bf  \Omega}|<1} \frac{1}{2}|\hat{\bm u}_{\bm k}|^2 $ is the
total energy in the Fourier plane perpendicular  to the rotation
direction.  In the inset of Fig.~\ref{bidim_dynamics} we show the
evolution of $E^{(2D)}(t)/E(t)$  as a function of time for four
different values of $Ro$.  It is clear that in both the split and the flux-loop
cascades,  the flow approaches asymptotically a quasi-2D  state, the
main difference is the time it takes to reach it. The main panel of
Fig.~\ref{bidim_dynamics} shows the time $t^*/\tau_f$ vs $Ro$ in a
log-lin plot at fixed $\lambda$,  where $\tau_f =
(\epsilon k_f^2)^{-1/3}$ is the characteristic time associated with the forcing. It is interesting to notice that when
entering the flux-loop region, $t^*$ increases with
$Ro$ faster than exponentially, indicating the presence of a possible
divergence at the critical value of $Ro_c\simeq 0.65$.  
The transition from the split to the forward cascade as a
function of the Rossby number has been also analyzed in a recent important 
study \cite{pestana_rossby-number_2020}, but with runs that evolved until
$t/\tau_f = 30$ only, and thus miss the development of the flux-loop 
regime.  In fact, the critical Rossby number reported in 
\cite{pestana_rossby-number_2020} is around the same value where we see the
transition from split to flux-loop cascades, and the growth rates they 
report (measured by the rate of change of the vertical correlations) depend
exponentially on $Ro$, the same way $t^*$ does for 
${Ro}<0.4$. 

Two  important remarks are now in order. First,  the transition between
split and flux-loop cascades exhibits hysteresis. If one takes a stable
simulation under the flux-loop regime and sufficiently  decreases
${Ro}$ the {vortex-crystal} is destabilized, the columns merge,
and  the system switches to a split cascade regime. Conversely, the
opposite does not happen. We have checked this explicitly. Second, and
even more important, the vortex crystals formed in the flux-loop regime
are metastable. For example, we observed that by evolving a simulation with
 parameters ($\lambda,Ro$) close to the transition from flux-loop to split cascade the vortex crystal structure is destabilized after a long time and the inverse cascade starts again (see the central panel in the SM movie). Metastability is consistent with
the fact that the vortex crystal is observed at the boundary between the
inverse and the forward cascade and with the existence of a critical
slowing down. 

{\sc Conclusions.}
By using huge high-performance-computing resources we have studied the
$(\lambda,Ro)$ phase space of rotating turbulence up to record resolution of
$512\times512\times32768$ grid points. We found the existence of a new
metastable flux-loop regime, where the inverse energy cascade is
stopped by a delicate balance between hetero- and homo-chiral triadic
non-linear interactions, leading to multiple metastable vortex-crystal
like states. These states are stable for very long times but can
transition to the inverse cascade regime if perturbed strongly enough.

Observations of multiple large-scale, self-organised turbulent state are
becoming more and more common in the turbulent literature, having been
observed in bounded flows \cite{huisman2014multiple},
in anisotropic sheared turbulence \cite{iyer2017multiscale}, in
swirling flows \cite{swirling} and in magnetohydrodamic flows
\cite{shebalin_broken_2010}.  They play a key role for both fundamental
aspects, suggesting the existence of multiple attractors in the system and
applied ones, leading to huge variations in the global energetic balance
for tiny changes in the control parameters.  The present study relates
these metastable states with the boundaries of two different states
(forward and split cascading) in a phase-diagram making the connection
with classical phase transitions.
 
Some important questions remain open, connected to the asymptotic behavior of the critical lines between 
different phases and the robustness of the metastable states in
the limit ${Ro} \to 0$, $\lambda \to \infty$ and $\lambda \to 0$
(see dashed red lines in the phase-space summary of Fig.
\ref{spectra}(a)). The presence of the flux-loop condensate prevented us
from extracting a precise functional behaviour in the ($\lambda,{Ro}$) plane
for the transition from forward to split cascade.  Although the
asymptotic scaling $\lambda \propto {Ro}^{-1}$ that was suggested
in \cite{alexakis2018cascades} and found in \cite{van2019critical} is
still plausible, the present results can not confirm nor reject this
hypothesis, because of the presence of the flux-loop phase.
Finally, we note that metastable properties can depend on the horizontal domain size $L$. If $L$ is large 
enough and the size of the
vortex-crystal is increased the possibility of having a destabilizing local defect increases. 

\begin{acknowledgments}
The authors acknowledge partial funding from the European Research
Council under the European Community's Seventh Framework Program, ERC
Grant Agreement No.~339032. The simulations have been done using resources provided through the PRACE initiative Pra$17\_4374$ at CINECA. The authors acknowledge Pablo Mininni and
Gregory Eyink for useful discussions. 
\end{acknowledgments}

\bibliography{biblio}


\clearpage
\onecolumngrid

\section{Supplemental Material}
\subsection{Numerical Simulations}
\label{sec:dns}
We performed a series of  direct numerical simulations (DNS) of the
incompressible Navier-Stokes equations (see Eqs.~\eqref{NSR} in the main text) where the normal viscosity has been
replaced with hyper-viscous dissipation: 
\begin{equation}
\label{eq:hy-navierstokes}
\begin{cases}
\partial_t \bm{u} + \bm{u} \cdot \nabla \bm{u} +2{\bm \Omega} \times {\bm u}
= - \nabla p + \nu (-1)^{\alpha + 1}\Delta^{\alpha}\bm{u} + \bm{f} \\
\nabla \cdot \bm{u} = 0,
\end{cases}
\end{equation}
with $\alpha=2$.  The motivation to use hyper-viscosity comes from request to minimize viscous effects at the forcing scale  at the expense of generating spectral bottlenecks
at high wavenumbers \cite{Falkovich94}, allowing the inverse energy cascade to develop with minimal Reynolds effects. For a few cases, we made sure to test that increasing the Reynolds numbers do not introduce important differences with what reported here. 
%
In Table~\ref{tbl:simulations} we provide a
list of all parameters used in each of the simulations presented in
this paper.
\begin{table}
\begin{center}
\begin{tabular}{|ccccc|}
$\lambda$ & $N_z$ & $\Omega$ & $Ro$ & Regime\\
\hline
1 & 512 & 37.5 & 0.204 & Split\\
1 & 512 & 34 & 0.225 & Split\\
1 & 512 & 33 & 0.232 & Split\\
1 & 512 & 30 & 0.255 & Split\\
1 & 512 & 26 & 0.294 & Split\\
1 & 512 & 22.5 & 0.340 & Split\\
1 & 512 & 20 & 0.382 & Split\\
1 & 512 & 18.75 & 0.408 & Split\\
1 & 512 & 17 & 0.450 & Flux-loop\\
1 & 512 & 15 & 0.510 & Flux-loop\\
1 & 512 & 15 & 0.501 & Flux-loop\\
1 & 512 & 13.5 & 0.566 & Flux-loop\\
1 & 512 & 12.5 & 0.611 & Flux-loop\\
1 & 512 & 11.6 & 0.659 & Forward\\
1 & 512 & 10.8 & 0.708 & Forward\\
1 & 512 & 10 & 0.764 & Forward\\
1 & 512 & 7.5 & 1.019 & Forward\\
2 & 1024 & 40 & 0.191 & Split\\
2 & 1024 & 32 & 0.239 & Split\\
2 & 1024 & 30 & 0.242 & Split\\
2 & 1024 & 28 & 0.274 & Split\\
2 & 1024 & 15 & 0.481 & Split\\
2 & 1024 & 27.5 & 0.278 & Flux-loop\\
2 & 1024 & 26.4 & 0.290 & Flux-loop\\
\end{tabular}
\begin{tabular}{|ccccc|}
$\lambda$ & $N_z$ & $\Omega$ & $Ro$ & Regime\\
\hline
2 & 1024 & 22.5 & 0.340 & Flux-loop\\
2 & 1024 & 18 & 0.425 & Flux-loop\\
2 & 1024 & 16.5 & 0.464 & Flux-loop\\
2 & 1024 & 14.5 & 0.528 & Flux-loop\\
2 & 1024 & 13.9 & 0.551 & Flux-loop\\
2 & 1024 & 12.5 & 0.613 & Forward\\
4 & 2048 & 70 & 0.109 & Split\\
4 & 2048 & 60 & 0.128 & Split\\
4 & 2048 & 50 & 0.153 & Flux-loop\\
4 & 2048 & 42.5 & 0.180 & Flux-loop\\
4 & 2048 & 30 & 0.249 & Flux-loop\\
4 & 2048 & 27.5 & 0.279 & Flux-loop\\
4 & 2048 & 20 & 0.383 & Flux-loop\\
4 & 2048 & 15 & 0.507 & Flux-loop\\
4 & 2048 & 13.9 & 0.551 & Forward\\
4 & 2048 & 12.5 & 0.613 & Forward\\
8 & 4096 & 30 & 0.246 & Flux-loop\\
16 & 8192 & 30 & 0.246 & Flux-loop\\
32 & 16384 & 30 & 0.244 & Flux-loop\\
64 & 32768 & 30 & 0.247 & Flux-loop\\
8 & 4096 & 15 & 0.511 & Forward\\
16 & 8192 & 15 & 0.480 & Forward\\
32 & 16384 & 15 & 0.481 & Forward\\
64 & 32768 & 15 & 0.479 & Forward\\
\end{tabular}
\end{center}
\caption{All simulations
were produced with a horizontal resolution of $N_x=N_y=512$, box size $L
= 2\pi$, hyperviscosity of order $\alpha=2$, forcing wavenumber in the
range range $[k_f, k_f+2]$ with $k_f=20$, forcing amplitude $f_0 =
f_1/\lambda$ with $f_0 = 1.66$, viscosity $\nu=4\times10^{-7}$, energy
injection rate of $\epsilon = 2.02$ and Reynolds number defined on the forcing scale 
${Re}=125$ (see main text). The parameters shown in the table are the box aspect
ratio $\lambda=H/L$, the number of collocation points in the vertical
direction $N_z$, rotation rate $\Omega$, Rossby number defined in terms
of the energy injection properties, $Ro = (\varepsilon_f
k_f^2)^{1/3}/\Omega$, and the regime the simulation is in.}
\label{tbl:simulations}
\end{table}

\subsection{Helical decomposition}
\label{sec:hel-dec}

In this section we provide the definition of the helical-fluxes calculated
in Fig.2(a-c) of the main text. For this purpose we exploit the
decomposition of any incompressible 3D flow into helical modes proposed
by \cite{waleffe_nature_1992,constantin_beltrami_1988}. 
From the incompressibility assumption it follows that $\bu(\bx)$ is a
solenoidal vector field, hence its Fourier modes $\bhu(\bk)$ depends
only on two linearly independent degrees of freedom and we can decompose
the velocity field as follows

\begin{equation}
\bhu_{\bk}(t) = \bhu_{\bk}^+(t) + \bhu_{\bk}^-(t) = \hat u^+_{\bk}(t)
{\bf h_+}(\bk) + \hat u^-_{\bk}(t) {\bf h_-}(\bk) \ ,
\end{equation}
where ${\bf h}_{\pm}(\bm{k})$ are the orthogonal eigenmodes of the curl
operator,
hence each Fourier modes of the velocity field satisfies    

\begin{equation}
i\bk \times \bhu_{\bk}^{s_{\bk}}= s_{\bk} \bhu_{\bk}^{s_{\bk}} \ ,
\end{equation}
with $s_{\bk} = \pm$. The homo-chiral  energy fluxes is  made out of triads with the same chirality:
\begin{align}
  \Pi^{hom}_k  = & \Pi^{(+,+,+)}_k + \Pi^{(-,-,-)}_k\\
  \Pi^{(\pm,\pm,\pm)}_k =& - i\sum_{|{\bk}| \le k} \sum_{\bp+\bq=\bk}\hspace{-0.5em} (\hat \bu^{\pm}_{-\bk} \cdot \hat\bu^\pm_{\bp}) (\bk \cdot \hat \bu^\pm_{\bq})
\end{align}
while the heterochiral is given by all resulting triads with two Fourier modes of opposite chirality:
\begin{equation}
\Pi^{het}_k = \Pi_k-\Pi^{hom}_k \ ,
\label{eq:flux_helical_dec}
\end{equation}
where $\Pi_k$ is the total energy flux defined in the main text as

\begin{equation}
\label{eq:flusso}
\Pi_k=- i \sum_{|\bm{k}|\le k}   \sum_{\bm{p}+\bm{q}=\bm{k}} (\hat{{\bm
u}}_{-\bm{k}} \cdot \hat{{\bm u}}_{\bm{p}}) ({\bm k} \cdot \hat{{\bm
u}}_{\bm{q}}) \ .
\end{equation}

\subsection{Video material}
\label{sec:movie}
A visualization of an  inverse and a flux-loop cascade can be found in uploaded video, 
where we show  the volume
rendering of the time evolution of the vorticity component along the direction of
rotation axis, $\bw_z$, for three different simulations obtained with $(Ro,\lambda)$:  $(0.25,1)$ split cascade (left panel);  $(0.51,1)$ split/flux-loop (center panel); and  $(0.58,1)$ flux-loop (right panel).
Notice the metastable regime  shown in the center panel where the inverse cascade first stops in to a quasi vortex crystal state and then suddenly restarts, thanks to 3D vortex merging.
In the bottom row of the same video we present the  energy
spectra and total energy evolution for the same  three simulations.

\end{document}